\documentclass[prl,amsmath,showpacs,amsfonts,amssymb,numbers,twocolumn,floatfix]{revtex4}
\usepackage{graphicx}
\usepackage{bm}
\usepackage{color}
\usepackage{mathrsfs}
\usepackage{wrapfig}
\usepackage{floatflt}

\begin{document}
\bibliographystyle{apsrev}

\title{Strong correlations via constrained-pairing mean-field theory}
\author{Takashi~Tsuchimochi and Gustavo~E.~Scuseria}
\affiliation{Department of Chemistry, Rice University, Houston, Texas 77005, USA 
}
\begin{abstract}
We present a mean-field approach for accurately describing strong correlations via electron number fluctuations and pairings constrained to an active space. Electron number conservation is broken and correct only on average but both spin and spatial symmetries are preserved. Optimized natural orbitals and occupations are determined by diagonalization of a mean-field Hamiltonian. This constrained-pairing mean-field theory (CPMFT) yields a two-particle density matrix ansatz that exclusively describe strong correlations. We demonstrate CPMFT accuracy with applications to the metal-insulator transition of large hydrogen clusters and molecular dissociation curves.
\end{abstract}


\maketitle

The efficient and accurate description of strong (static or nondynamical) correlations in electronic structure theory has remained an elusive goal. 
Despite its importance in many physical and chemical processes, a first-principles, accurate, black-box, and computationally inexpensive scheme for static correlation remains unknown. 
We present in this Letter an interesting alternative that allows to treat strong correlations in large molecular systems and opens a way 
to treat extended systems too. 
We build upon previous work of Staroverov and Scuseria\cite{HFB2}, who showed that 
a Hartree-Fock-Bogoliubov (HFB)\cite{Blaizot,Ring} scheme with an effective pairing interaction of $-\frac{\zeta}{r_{12}}$ 
can accurately describe the strong correlations occurring 
in the 4-electron series and in molecular dissociations done over the correct spatial- and spin-symmetry $restricted$ HF (RHF) reference. 
For $\zeta=1$ (referred to as 1HFB below), this scheme is equivalent to the functional known as Corrected Hartree-Fock (CHF)\cite{CHF} in density matrix functional theory\cite{Gilbert}. 
1HFB allows for efficient optimization of natural orbitals (NOs) and occupations via matrix diagonalization of a mean-field Hamiltonian 
of small dimension (twice the number of orbitals). The HFB wavefunction breaks electron number conservation thus requiring a chemical potential 
to adjust the electron count to its correct value $N$. HFB is connected to the $N$-electron multi configurational antisymmetrized geminal 
power wavefunction\cite{HFB2} (known as pfaffian in recent literature\cite{pfaffian}), thus the appearance of static correlation 
should not be unexpected. 
For most molecules near equilibrium, 1HFB does not include correlations and yields the RHF solution. 
As the system dissociates, 1HFB adds strong (left-right) correlations but significantly overcorrelates at large 
internuclear separations in almost all cases. This behavior was considered 
a serious drawback\cite{DMFT4}, and therefore the model lost attention.
In this Letter, however, we show that by incorporating simple additional constraints, a mixture of HF and HFB methods for inactive 
and active sets of orbitals, respectively, can overcome the overcorrelation problem and yield an accurate mean-field approximation 
for strong correlations. Our model is here denoted constrained-pairing mean-field theory (CPMFT). To add important dynamical correlations 
near equilibrium, we propose a straightforward extension of the TPSS correlation functional\cite{TPSS}, 
which is applicable to the hydrogen clusters discussed below.
Throughout this Letter, we will use exclusively the real NO basis, \{$\varphi_i$\}, where the one-particle density matrix (1PDM) is diagonal.

{\it Theory.} We partition the orbitals into core, active, and virtual spaces. Only the orbitals in the active space are subject to a Bogoliubov transform; 
the core and virtual orbitals are treated by HF. In this way, we {\it constrain} pairing interactions exclusively to the active space. 
The CPMFT wavefunction is therefore a BCS ansatz\cite{Blaizot,Ring} mixed with HF
\begin{eqnarray}
|\Psi^{\rm CPMFT}\rangle=\prod_c a_{c}^\dag a_{\bar{c}}^\dag\prod_i (x_i+y_i a_{i}^\dag a_{\bar{i}}^\dag)|0\rangle .
\label{eq:CPMFTWF}
\end{eqnarray}
Here $|0\rangle$ indicates the vacuum, a bar on a suffix means down spin, and indices $c$ and $i$ run over core and active spaces, respectively. Normalization requires $x_i^2+y_i^2=1$ and $y_i^2=n_i$ are occupation numbers of the 1PDM $\gamma_{ij}=\langle a_i^\dag a_j\rangle$. Eq.(\ref{eq:CPMFTWF}) includes $N$-electron determinants as well as configurations with electron number different from $N$. Double excitations are included in the CPMFT ansatz. The pairing matrix $\kappa_{ij}=\langle a_i^\dag a_j^\dag \rangle $ is $\kappa^2 = \gamma-\gamma^2$ and is diagonal in the NO basis, $\kappa_i=x_iy_i = \sqrt{n_i(1-n_i)}$. 
Our formalism is closed-shell, {\it i.e.} $n_\alpha = n_\beta = n$ and $0 \le n \le 1$.
Note that $\kappa_i$ is 0 when $n=0 ~or ~1$ and  $\kappa_i$ is  $\frac{1}{2}$ when correlation is all static, {\it e.g.} 
at dissociation.  More details about HFB can be found in Refs.[\onlinecite{HFB2,Blaizot,Ring}]. 

In CPMFT, we introduce additional constraints to divide orbital spaces into fully occupied (core, $n=1$), fractionally occupied (active), and virtual ($n=0$), resulting in the energy expression
\begin{eqnarray}
E^{\rm CPMFT}&=&E^{\rm 1HFB} + \mu_{\rm C}\left(\sum_c n_c-N_{\rm C}\right)\nonumber\\
&&+\mu_{\rm V}\sum_v n_v+\mu_{\rm A}\left(\sum_i n_i-N_{\rm A}\right),
\label{eq:ECPMFT}
\end{eqnarray}
where $v$ runs over the virtual space, $N_{\rm C}$ and $N_{\rm A}$ are the the number of electrons in the core and active spaces, respectively, and the different $\mu$'s are the chemical potentials (Lagrange multipliers) constraining the electron numbers in each space. The 1HFB energy above is

\begin{eqnarray}
E^{\rm 1HFB}&=&2\sum_i n_i h_{ii}+\sum_{ij}\left[n_in_j (2J_{ij}-K_{ij})-\kappa_i\kappa_j K_{ij}\right],\nonumber\\
\label{eq:EHFB}
\end{eqnarray}
where $h_{ij}$, $J_{ij}=\langle \varphi_i\varphi_j|\varphi_i\varphi_j \rangle$, and $K_{ij}=\langle \varphi_i\varphi_j|\varphi_j\varphi_i \rangle$ are one-electron, two-electron Coulomb and exchange integrals, respectively.  The two-particle density matrix (2PDM) of this ansatz is 
\begin{eqnarray}
\Gamma_{ij}^{kl}=\frac{1}{2}\left(\gamma_{ik}\gamma_{jl}-\gamma_{il}\gamma_{jk}-	\kappa_{ij}\kappa_{kl}\right),
\label{eq:2PDM}
\end{eqnarray}
which includes particle-particle and hole-hole but no particle-hole correlation terms. As shown below, this simple ansatz describes strong correlations very accurately. We conjecture that the cumulant term ($\kappa \otimes \kappa$ with $\kappa^2 = \gamma - \gamma^2$) offers a compelling definition of what we intuitively understand as static or strong correlation. Note that the negative sign of the last term in Eq.(\ref{eq:2PDM}) comes from the attractive pairing potential.

The mathematical solution to the CPMFT constrained problem is to set $\mu_{\rm C}=-\infty$ and $\mu_{\rm V}=\infty$. The physical meaning of this is that $\mu_{\rm C}$ ($\mu_{\rm V}$) pulls down (pushes up) the energies of core (virtual) orbitals so that these levels do not mix with the active orbitals. In our calculations, we set the chemical potential for the inactive spaces $\mu_{\rm V} = -\mu_{\rm C}$, and gradually increase $\mu_{\rm V}$ smoothly to convergence.

As a benchmark example, we discuss first the dissociation of H$_2$, a paradigmatic example of strong correlation. The CPMFT(2,2) (two electrons in two active orbitals) wavefunction for such a system is given by
\begin{eqnarray}
|\Psi^{\rm CPMFT}\rangle&=&\frac{1}{2}\left(|0\rangle+|\varphi_{\sigma_{\rm 1s}}\bar{\varphi}_{\sigma_{\rm 1s}}\rangle+|\varphi_{\sigma^*_{\rm 1s}}\bar{\varphi}_{\sigma^*_{\rm 1s}}\rangle\right.\nonumber\\
&&\left.+|\varphi_{\sigma_{\rm 1s}}\bar{\varphi}_{\sigma_{\rm 1s}}\varphi_{\sigma^*_{\rm 1s}}\bar{\varphi}_{\sigma^*_{\rm 1s}}\rangle\right).
\label{eq:1HFBWF}
\end{eqnarray}
Note that the first and last determinants of Eq.(\ref{eq:1HFBWF}) contains 0 and 4 electrons, respectively. However, the expectation value of the electron number operator on this CPMFT wavefunction is 2. The energy associated with Eq.(\ref{eq:1HFBWF}) is
\begin{eqnarray}
\label{eq:E1HFBHF}
E^{\rm CPMFT}&=&h_{\sigma_{\rm 1s}\sigma_{\rm 1s}}  + h_{\sigma^*_{\rm 1s}\sigma^*_{\rm 1s}} +J_{\sigma_{\rm 1s}\sigma^*_{\rm 1s}}-K_{\sigma_{\rm 1s}\sigma^*_{\rm 1s}},
\end{eqnarray}
and given that at dissociation, $J_{\sigma_{\rm 1s}\sigma^*_{\rm 1s}}=K_{\sigma_{\rm 1s}\sigma^*_{\rm 1s}}$, this expression correctly reproduces the energy of two isolated hydrogen atoms. Eq.(\ref{eq:1HFBWF}) also yields the exact 1PDM at dissociation.

\begin{figure}[t]
  \includegraphics[width=80mm]{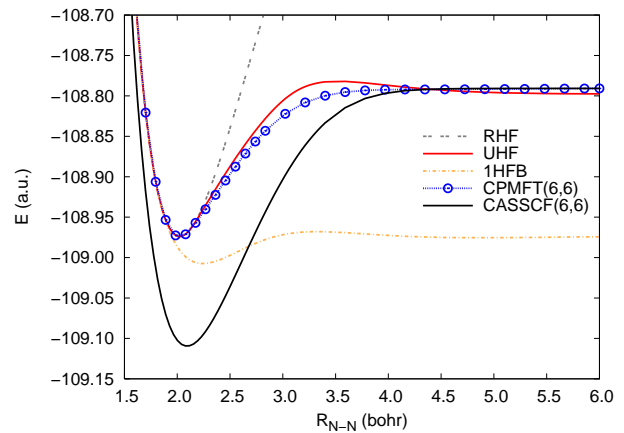}
 \caption{Potential energy curves of N$_2$ calculated with a Gaussian 6-311++G** atomic orbital basis set.}
 \label{fig:FIG1}
\end{figure}

{\it Results.} We have implemented CPMFT in the {\sc Gaussian} suite of programs\cite{Gaussian}. As another paradigmatic example of strong correlation, we present results for the dissociation potential of the N$_2$ molecule in Fig.~\ref{fig:FIG1}. This is a challenging case because it requires up to six-electron excitations to yield the correct curve. CPMFT(6,6) gives a very accurate description of this dissociation at low (mean-field) computational cost.
It accomplishes this feat while preserving spatial and spin symmetries. Unrestricted HF (UHF) improves over RHF, but the density is symmetry-broken with three $\alpha$ electrons localizing on one N atom, and three $\beta$ electrons on the other. Complete Active Space Self-Consistent Field CASSCF(6,6) calculations \cite{Jensen}, which include all excitations for 6 electrons within 6 active orbitals, also yield a correct potential curve for N$_2$ including a part of dynamic and static correlations near equilibrium and only strong correlations towards dissociation.
Note how 1HFB overcorrelates over the entire curve yielding very poor results.

While CPMFT significantly improves over HF and 1HFB in diatomics, we would like to present here its application to large systems that are currently completely out of the reach of other approaches for accurately describing electron correlations. Accordingly, we will investigate below large hydrogen networks. Before doing so, it should be evident from Fig.~\ref{fig:FIG1} that CPMFT lacks dynamical correlation, especially at equilibrium, where the total correlation energy is underestimated. 
Density functional theory (DFT) seems a natural candidate for adding this effect. However, it is known\cite{Savin88,Discon} that semilocal DFT correlation functionals face substantial challenges when applied to symmetry adapted densities as those resulting from CPMFT (as opposed to symmetry broken unrestricted models). 
Alternatives to circumvent this issue have been proposed in the literature\cite{Savin88,Perdew95}. 
In order to deal with this problem and with large hydrogen clusters in mind, we here propose a simple solution that uses $\kappa$ as a ``detector" of static correlation.
We introduce a correction factor to DFT correlation functionals using $g_i=1-\left(2\kappa_i\right)^m$ and define $g({\bf r}) = \sum_i g_i n_i |\varphi_i({\bf r})|^2$, where $m$ is a parameter controlling the rate of change of $g_i$. We thus write
\begin{eqnarray}
E_{\rm c}^{\rm \kappa}=\int \epsilon_{\rm c}({\bf r})  g({\bf r}) d{\bf r},
\end{eqnarray}
where $\epsilon_{\rm c}({\bf r})$ is the DFT correlation energy density. We have modified the TPSS correlation functional to include $ g({\bf r}) $ and define $E_{\rm c}^{\rm \kappa TPSS}$. 

\begin{figure}[t]
  \includegraphics[width=80mm]{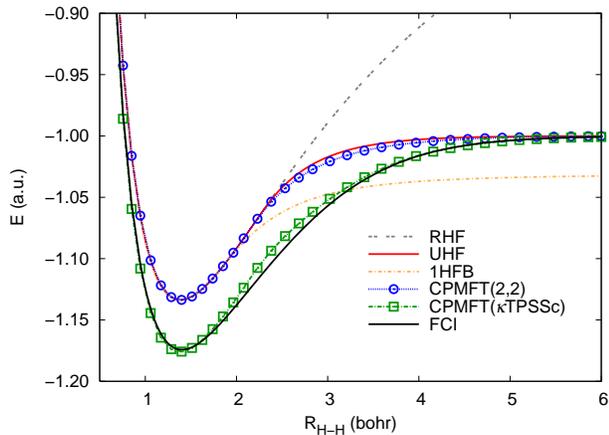}
 \caption{Potential energy curves for H$_2$ dissociation calculated with a Gaussian cc-$p$V5Z basis set.}
 \label{fig:FIG2}
\end{figure}

In Fig.~\ref{fig:FIG2}, we plot the potential energy curve of H$_2$. For R$_{\rm H-H}\rightarrow\infty$, correlation in this system is all static while near equilibrium, there is substantial dynamical correlation. Here, $E_{\rm c}^{\rm \kappa TPSS}$ is evaluated with the CPMFT density and NOs, hence our calculations are not self-consistent, and we shall denote them as CPMFT($\kappa$TPSSc). The parameter $m$ is chosen as 10 throughout this Letter based on a fit of CPMFT($\kappa$TPSSc) to the the exact full configuration interaction (FCI)\cite{Jensen} dissociation curve of H$_2$.

We now discuss our results on hydrogen networks, namely, a one dimensional (1D) H$_{50}$ chain and a three dimensional (3D) $6\times 6\times 6$ hydrogen cube, with an internuclear distance R$_{\rm H-H}$ between adjacent atoms. In the limit of R$_{\rm H-H}\rightarrow \infty$, these model systems dissociate to 50 and 216 isolated hydrogen atoms, respectively. As R$_{{\rm H-H}}$ increases, they display a metal-insulator transition. Both systems are paradigmatic models for strongly correlated Mott insulators\cite{DMRG2} and are not treatable by CASSCF due to the staggering number of active configurations, $10^{27}$ and $10^{123}$ for the 1D and 3D cases, respectively. In the linear case, the density matrix renormalization group (DMRG)\cite{DMRG1} approach, a very accurate wavefunction method, is available\cite{DMRG2}. However, for the 3D case, DMRG is not (yet) applicable. 
As shown below, accurate CPMFT of these two model systems only requires diagonalizing Hamiltonians of moderate dimension, $100\times 100$ and $432\times 432$, respectively.

\begin{figure}[t]
  \includegraphics[width=80mm]{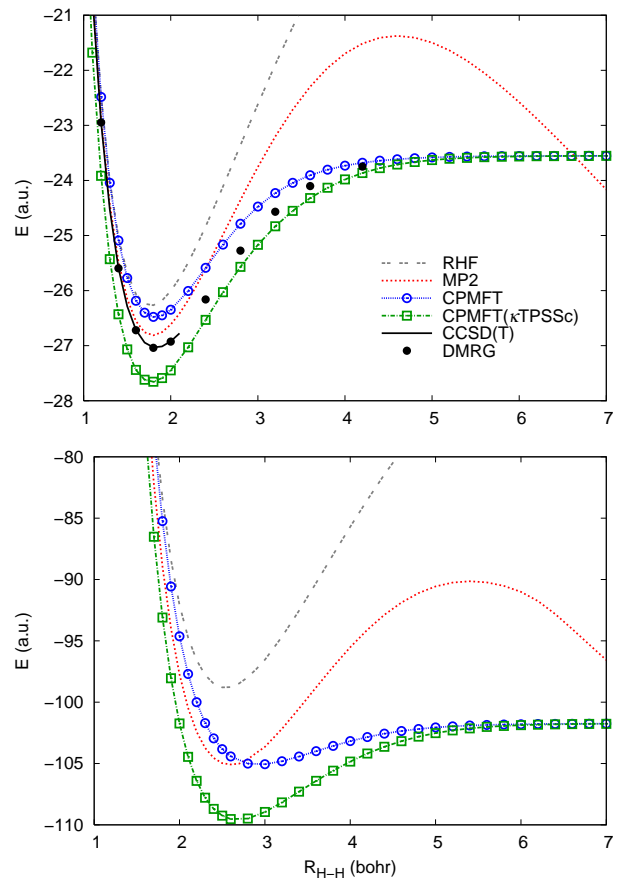}
 \caption{({\it Top}) Potential energy curves for the symmetric dissociation of a H$_{50}$ chain. DMRG results are taken from Ref.[\onlinecite{DMRG2}]. ({\it Bottom}) Same as Top, but for a $6\times 6\times 6$ hydrogen cube.}
 \label{fig:FIG3}
\end{figure}

Fig.~\ref{fig:FIG3} presents potential energy curves for each system computed with several methods and an STO-6G basis. This choice of minimal basis is made to allow comparison with results in the literature\cite{DMRG2}. All electrons and orbitals are explicitly treated in our calculations. The active spaces of CPMFT are (50,50) and (216,216) for the 1D and 3D systems, respectively. This means that for these clusters $N_C=0$ and $N_A$ is equal to the size of the active space in Eq.(\ref{eq:ECPMFT}).
In both cases, RHF and second-order M{$\o$}ller-Plesset perturbation theory (MP2)\cite{Jensen} yield unreliable curves, while coupled cluster singles, doubles, and perturbative triples (CCSD(T))\cite{Jensen} has convergence difficulties at long internuclear distances and hence is not included in the plot for the 3D case.
The RHF method misses a considerable amount of strong correlation. MP2 may both miss or
exaggerate non-dynamical correlation, as goes forth from Fig. 3. CCSD(T), despite its
single reference character, is quite good at covering non-dynamical
correlation, thanks to its inclusion of infinite-order terms in
a balanced way. If there are problems for long R values, they are
not visible in Fig. 3 due to the mentioned convergence problems. 
For the 1D system around equilibrium, the DMRG energy is very similar to that of CCSD(T). While CCSD(T) breaks down past the equilibrium bond length, DMRG is able to predict accurately the potential curve toward dissociation. 
\begin{table}[t]
\caption{Correlation energy (a.u.) of a H$_{50}$ chain at R$_{\rm e}$.}
\begin{tabular*}{8.5cm}{@{\extracolsep{\fill}}ccccc}\hline\hline
Basis set&MP2&CCSD(T)&CPMFT&CPMFT($\kappa$TPSSc)\\
\hline
STO-6G\footnote[1]{The DMRG correlation energy with STO-6G basis is -0.772 67 (Ref.[\onlinecite{DMRG2}]).}  &-0.545 55  &-0.765 47 &-0.211 12 & -1.384 31\\
cc-$p$VDZ&-0.976 45 &-1.213 41 &-0.203 42&-1.426 27\\
cc-$p$VTZ&-1.123 26 &-1.347 9\footnote[2]{Estimated value by extrapolaion of a H$_n$ chain to $n=50$.} &-0.205 46 &-1.408 85 \\
cc-$p$VQZ&-1.162 74 &-1.374\footnotemark[2]& -0.205 81&-1.410 06 \\
\hline
\hline
\end{tabular*}
\label{tb:CPMFTLYP}
\end{table}

On the other hand, CPMFT and CPMFT($\kappa$TPSSc) yield correct dissociation curves (within the basis set used)  not only for the 1D system but also for the 3D case.  Only CPMFT properly describes the metal-insulator transition of the 3D hydrogen network. Although at first glance CPMFT($\kappa$TPSSc) appears to overcorrelate the 1D chain near equilibrium, this is partially a basis set effect: in the top panel of Fig.~\ref{fig:FIG3} CCSD(T) is essentially underestimating correlation effects due to the small basis used in the calculation (a necessity to make the CCSD(T) calculation affordable). 
\begin{figure}[t!]
  \includegraphics[width=80mm]{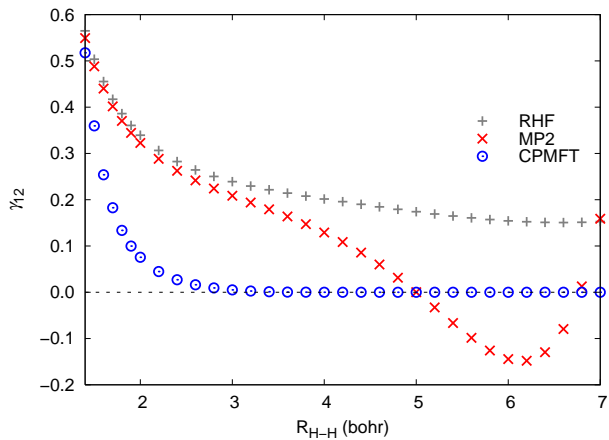}
 \caption{Decay of off-diagonal density matrix term ($\gamma_{12}$) for two hydrogen atoms at diagonal vertices in a $4\times 4\times 4$ hydrogen cube.}
 \label{fig:C-C}
\end{figure}

To further investigate the accuracy of CPMFT, we have carried out calculations of the 1D hydrogen chain near equilibrium (R$_{\rm e}=1.8$ bohr) with very large cc-$p$V$n$Z basis sets, where $n={\rm D,T,Q}$. We compare our correlation energies with CCSD(T), which yields results close to DMRG (Fig.~\ref{fig:FIG3}) with the STO-6G basis and thus we consider it very accurate for the 1D chain near R$_{\rm e}$. For the cc-$p$VTZ and cc-$p$VQZ bases, we extrapolated the CCSD(T) correlation energy of H$_{50}$ from shorter chains, as it was not feasible to compute CCSD(T) for H$_{50}$ within reasonable time. Results are presented in TABLE~\ref{tb:CPMFTLYP}. While convergence of the CCSD(T) correlation energy with basis set size is slow, the correlation energies of both CPMFT and CPMFT($\kappa$TPSS) converge reasonably well. Static correlation is expected to be fairly basis set independent, a property reproduced by CPMFT.

In order to visualize the metal-insulator transition explicitly, we have plotted the off-diagonal density matrix term $\gamma_{12}$ between two hydrogen atoms at diagonal vertices in a $4 \times 4 \times 4$ cube (Fig.~\ref{fig:C-C}). Both RHF and MP2 off-diagonal terms remain substantially above zero as R$_{\rm H-H}$ increases, implying  that the electrons are delocalized. The CPMFT off-diagonal $\gamma_{12}$, on the other hand, rapidly decays toward zero. Evidently, only CPMFT reveals the gradual metal-insulator transition in this hydrogen cube.

Finally, we would like to mention some current limitations and prospects of our CPMFT model. When applied to hetero-nuclear systems, CPMFT does not reproduce the exact dissociation energy although there is a very substantial improvement over RHF and 1HFB. 
For example, the CPMFT dissociation energy error relative to the correct CASSCF limit is 0.2 eV for BH with a 6-311++G(3df,3pd) Gaussian basis set. Ways to address and correct dissociations in hetero-nuclear systems to non-degenerate orbitals are currently being investigated and will be discussed in future publications. A more elaborate scheme for dealing with dynamical correlations for symmetry adapted densities is also being developed. Most importantly, we would like to stress our conjecture that the particular CPMFT form of the 2PDM in Eq.(\ref{eq:2PDM}) is most adept for describing strong correlations.

We thank Tom Henderson and Andreas Savin for helpful discussions. This work was supported by NSF CHE-0807194 and the Welch Foundation (C-0036).

\end{document}